\documentclass[letterpaper]{article} 
\usepackage{aaai2026}  
\usepackage{times}  
\usepackage{helvet}  
\usepackage{courier}  
\usepackage[hyphens]{url}  
\usepackage{graphicx} 
\usepackage{natbib}  
\usepackage{caption} 
\usepackage{algorithm}
\usepackage{algorithmic}
\usepackage{amsmath}
\usepackage{longtable}
\usepackage{array}
\usepackage{makecell}
\usepackage{xpatch}
\usepackage{multirow}
\usepackage{bbding}
\usepackage{booktabs}
\usepackage{amsfonts}

\usepackage{newfloat}
\usepackage{listings}
\DeclareCaptionStyle{ruled}{labelfont=normalfont,labelsep=colon,strut=off} 
\floatstyle{ruled}
\newfloat{listing}{tb}{lst}{}
\floatname{listing}{Listing}
\title{S-DAG: A Subject-Based Directed Acyclic Graph for Multi-Agent Heterogeneous Reasoning}
\author{
    Jiangwen Dong\equalcontrib\textsuperscript{\rm 1}, Zehui Lin\equalcontrib\textsuperscript{\rm 1}, Wanyu Lin\thanks{Corresponding Author: Wanyu Lin, wan-yu.lin@polyu.edu.hk}\textsuperscript{\rm 1, 2}, Mingjin Zhang\textsuperscript{\rm 2}
}
\affiliations{
    \textsuperscript{\rm 1}Department of Data Science and Artificial Intelligence, The Hong Kong Polytechnic University, Hong Kong SAR, China\\
    
    \textsuperscript{\rm 2}Department of Computing, The Hong Kong Polytechnic University, Hong Kong SAR, China\\

}

\begin{document}

\maketitle

\begin{abstract}
Large Language Models (LLMs) have achieved impressive performance in complex reasoning problems. Their effectiveness highly depends on the specific nature of the task, especially the required domain knowledge. Existing approaches, such as mixture-of-experts, typically operate at the task level; they are too coarse to effectively solve the heterogeneous problems involving multiple subjects. This work proposes a novel framework that performs fine-grained analysis at the subject level equipped with a designated multi-agent collaboration strategy for addressing heterogeneous problem reasoning. Specifically, given an input query, we first employ a Graph Neural Network to identify the relevant subjects and infer their interdependencies to generate an~\textit{Subject-based Directed Acyclic Graph} (S-DAG), where nodes represent subjects and edges encode information flow. Then, we profile the LLM models by assigning each model a subject-specific expertise score, and select the top-performing one for matching the corresponding subject of the S-DAG. Such subject-model matching enables graph-structured multi-agent collaboration where information flows from the starting model to the ending model over S-DAG. We curate and release multi-subject subsets of standard benchmarks (MMLU-Pro, GPQA, MedMCQA) to better reflect complex, real-world reasoning tasks. Extensive experiments show that our approach significantly outperforms existing task-level model selection and multi-agent collaboration baselines in accuracy and efficiency. These results highlight the effectiveness of subject-aware reasoning and structured collaboration in addressing complex and multi-subject problems.
\end{abstract}

\begin{links}
    \link{Code}{https://github.com/WanyuGroup/AAAI2026_S-DAG}
    \link{Extended version}{https://arxiv.org/abs/2511.06727}
\end{links}

\section{Introduction}
\label{sec:intro}

In recent years, intelligent agents based on large language models (LLMs) have developed rapidly and achieved significant advancements across various fields, ranging from question answering~\cite{yue2025survey,zhuang2023toolqa} to text generation~\cite{huang2023grounded, wu2024autogen} and complex reasoning tasks~\cite{ke2025survey,zhang2024llm}.
While the development of a general-purpose LLM is promising~\cite{mumuni2025large,kojima2022large}, it becomes evident that a single LLM often struggles to handle complex reasoning tasks, especially when these problems span multiple disciplines~\cite{feng2025one}. This limitation raises higher demands for model training and fine-tuning~\cite{hoffmann2022training}. In this context, multi-agent systems based on LLMs have emerged, aiming to leverage the collective intelligence and specialized expertise of multiple agents to tackle complex and multidisciplinary problems~\cite{guo2024large,du2023improving,talebirad2023multi,han2024llm,gu2025explain, liang2024encouraging,yao2025socialized}\footnote{For simplicity, agents, LLMs, and models are used interchangeably.}.

\begin{figure}[!t]
    \centering \includegraphics[width=\linewidth]{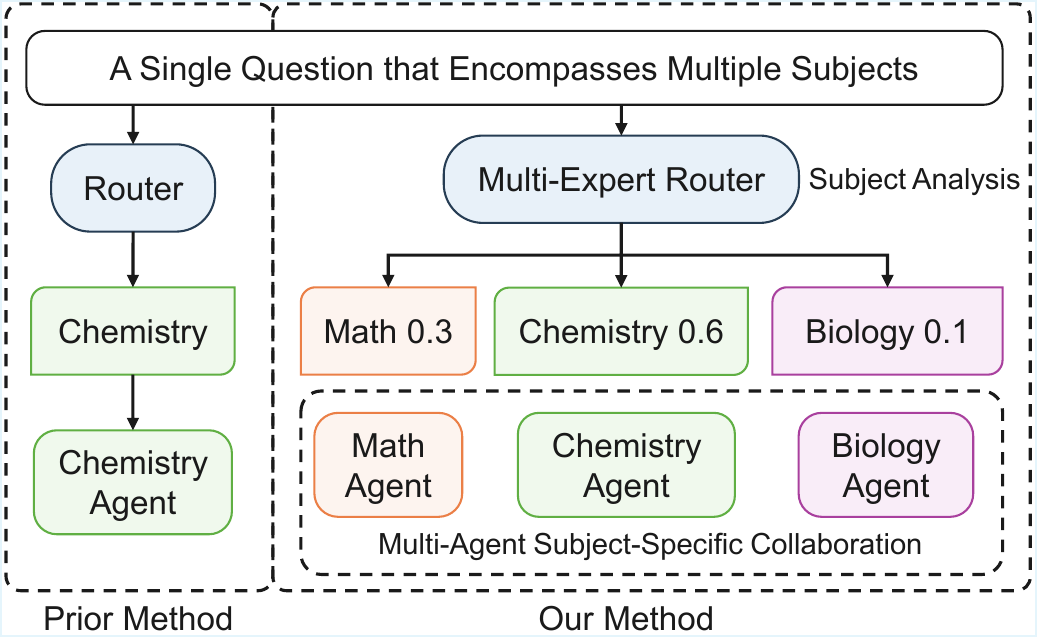}
    \caption{Comparison of the prior method with single agent and the proposed S-DAG approach with multi-agent collaboration. The prior method routes the problem to a single agent based on a coarse domain label, while our S-DAG approach conducts fine-grained subject analysis, identifying multiple relevant domains with associated relevant weights.}
    \label{fig:comparison}
\end{figure}

Existing research has explored the mixture-of-experts (MoE) framework that dynamically selects the most suitable LLMs for a given problem~\cite{masoudnia2014mixture,zhou2022mixture,cai2024survey}. Subsequently, the mixture-of-agents (MoA) paradigm leveraging multi-agent collaboration is proposed to deal with more complex problems by combining the strengths of diverse LLMs~\cite{du2023improving,zhang2024chain, wang2024mixture,li2024smoa}. The most relevant to us is Symbolic-MoE, which analyzes the required subject knowledge for a task and then utilizes a set of top-k expert/subject models to solve the heterogeneous problem~\cite{chen2025symbolic}. For clarity, we organize the existing heterogeneous reasoning paradigms in Table~\ref{tab:comparison}. These prior works often assume queries belong to a single knowledge domain or simply rely on a single ``best'' model or agent for reasoning~\cite{chen2025symbolic,feng2024knowledge,feng2025heterogeneous,feng2025graphrouter}. Very rare work considers the fine-grained subject-specific information of the problem, not to mention multi-agent collaboration at the subject level, as shown in Figure \ref{fig:comparison}. Such limitation hinders their applicability of prior works in heterogeneous reasoning tasks, where seamless integration of cross-domain knowledge is critical. Therefore, this paper aims to address the following research problem: \textit{How can we optimally select and coordinate expert LLMs at subject level for complex, multi-subject problems to achieve both high accuracy and efficiency?}

\begin{table*}[!t]
\small
\centering
\scalebox{0.85}{
\begin{tabular}{c c c c c}
\toprule
Category & Example Work & Subject-Level Analysis & MAS Collaboration & Subject-Specific Collaboration \\ \midrule
\multirow{2}{*}{Routing} & GraphRouter~\cite{feng2025graphrouter} & \XSolidBrush & \XSolidBrush & \XSolidBrush \\
 & SymbolicMoE~\cite{chen2025symbolic} & \Checkmark & \XSolidBrush & \XSolidBrush \\ \midrule
\multirow{2}{*}{Multi-Agent System (MAS)} & Heterogeneous Swarm~\cite{feng2025heterogeneous} & \XSolidBrush & \Checkmark & \XSolidBrush \\
 & Knowledge Card~\cite{feng2024knowledge} & \Checkmark & \XSolidBrush & \XSolidBrush \\ \midrule
Routing for MAS & S-DAG (Ours) & \Checkmark & \Checkmark & \Checkmark \\
\bottomrule
\end{tabular}}
\caption{Comparison of our proposed S-DAG and prior methods for heterogeneous reasoning. Unlike prior methods that either perform single-domain routing or lack fine-grained problem understanding, S-DAG supports detailed subject-level analysis and enables dynamic, subject-specific multi-agent collaboration for more effective reasoning.}
\label{tab:comparison}
\end{table*}

\begin{figure*}[t!]
    \small
    \centering
    \includegraphics[width=0.93\linewidth]{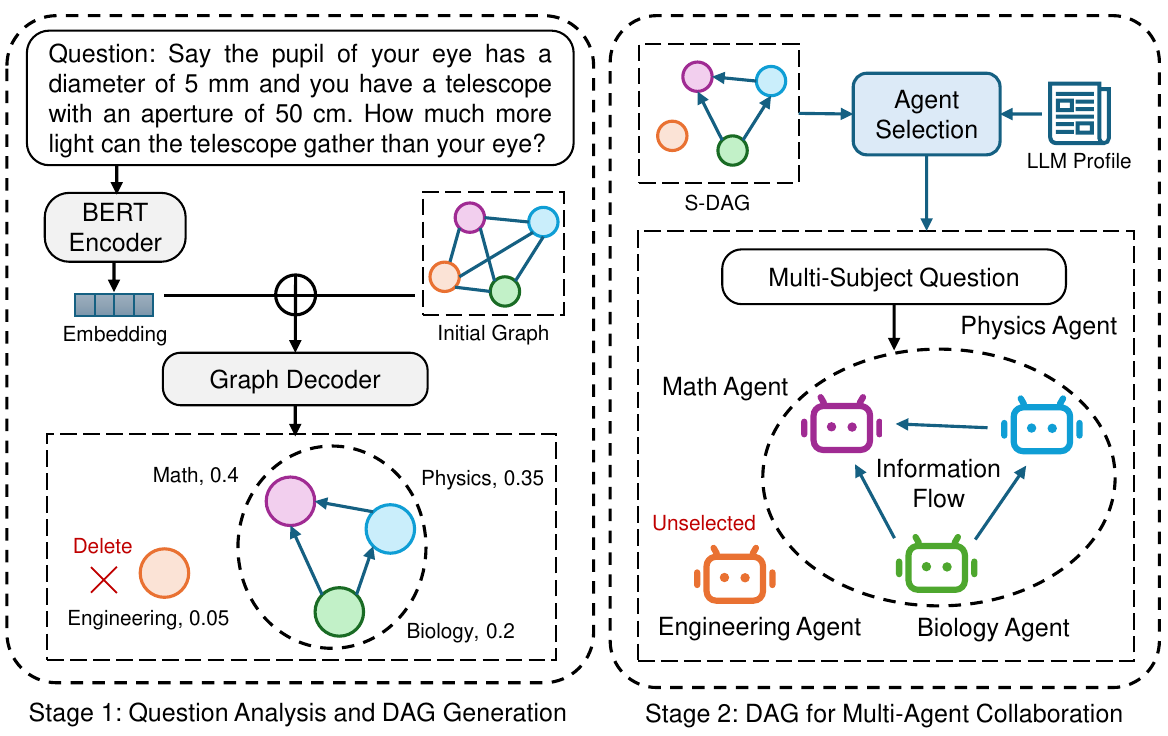}
    \caption{The Overview of the S-DAG Framework. The framework operates in two stages. In Stage 1, the input question is encoded using a BERT encoder, and a Graph Decoder generates the S-DAG, capturing subject dependencies and pruning irrelevant subjects. In Stage 2, expert LLMs are selected based on their subject-specific expertise and organized according to the S-DAG, with directed edges defining the information flow for multi-subject collaborative reasoning.}
    \label{fig:method}
\end{figure*}

In this work, we propose Subject-based Directed Acyclic Graph (S-DAG), as shown in Figure \ref{fig:method}, a novel framework for addressing heterogeneous reasoning problems that require knowledge across multiple subject domains. The S-DAG identifies the relevant subjects for a given problem and defines the graph-structured information flow for multi-agent collaboration. We begin by modeling the complete set of subjects as a fully connected graph. To extract fine-grained subject-level structure, we introduce a specialized graph neural network that learns node embeddings to capture the relevant subjects and their interdependencies with respect to a given problem. From this, we derive a subject-based directed acyclic graph that reflects the essential subjects and reasoning flow for the problem. Based on the constructed S-DAG, we perform subject–LLM matching by profiling LLMs according to their subject-specific capabilities. The constructed S-DAG and LLM profile guide a structured multi-agent collaboration mechanism, where domain-specialized LLMs are assigned to subject nodes and communicate according to the DAG topology. Through this design, our S-DAG enables efficient and subject-level multi-agent reasoning. In summary, the main contributions of this work are as follows:

\begin{itemize}
    \item To the best of our knowledge, we are the first to study the heterogeneous reasoning problem that a single complex problem covers multiple subject knowledge. Our novel framework, S-DAG, enables fine-grained subject-specific decomposition and graph-structured multi-agent collaboration mechanism for the multi-subject problem.
    
    \item We develop a fine-grained subject–LLM matching strategy by profiling LLMs according to subject-specific capabilities, enabling precise assignment of expert agents and efficient coordination via the S-DAG.

    \item We curate multi-subject evaluation datasets by manually selecting samples that require multi-subject knowledge from three challenging benchmarks—MMLU-Pro, GPQA, and MedMCQA. Extensive experiments demonstrate that our approach substantially outperforms both single-model and multi-model baselines in terms of accuracy and computational efficiency.
\end{itemize}

\section{Related Work}

\textbf{Multi-Agent Systems.}
Recent advances in Multi-Agent Systems (MAS) have introduced diverse collaboration mechanisms to tackle complex tasks~\cite{guo2024large,talebirad2023multi,han2024llm}, broadly categorized into fixed and dynamic paradigms. (1) \textit{Fixed Multi-Agent Systems} rely on manually designed architectures, such as LLM debates~\cite{du2023improving,liang2024encouraging}, chain-of-agents~\cite{gu2025explain,zhang2024chain,tao2025chain}, and graph-based systems~\cite{yin2023exchange,li2024improving,lin2025graph}. These approaches enable collaboration among agents with predefined roles and structures, making them effective for well-structured problems, but often lacking adaptability to dynamic tasks. (2) \textit{Dynamic Multi-Agent Systems} adapt their structure in response to real-time task demands. Notable examples include GPTSwarm~\cite{zhuge2024gptswarm}, GHG~\cite{li2025graphs} and Heterogeneous Swarms~\cite{feng2025heterogeneous}, which optimize collaboration through dynamic graph structures. AgentPrune~\cite{zhang2024cut} improves communication efficiency by pruning redundant links, while DyLAN~\cite{liu2024dynamic} dynamically selects agents and communication paths based on task context. MasRouter~\cite{yue2025masrouter} introduces a cascaded controller for mode selection, role assignment, and LLM routing, enabling efficient and adaptive MAS construction.

\textbf{Heterogeneous Reasoning.}
Heterogeneous reasoning focuses on solving problems that require knowledge across multiple domains~\cite{xin2024atomr,chen2025symbolic,feng2024knowledge}. Existing approaches can be grouped by the number of models involved: (1) \textit{Single-Model Approaches} select one expert model per query, often using routing mechanisms. MoE techniques~\cite{masoudnia2014mixture,zhou2022mixture,cai2024survey} specialize models over input space, while FrugalGPT~\cite{chen2024frugalgpt} uses a reliability predictor, and GraphRouter~\cite{feng2025graphrouter} employs a graph neural network to frame selection as edge prediction. Though effective, these methods fall short when queries require multi-domain expertise. (2) \textit{Multi-Model Approaches} enable collaboration across multiple expert models. Mixture-of-Agents (MoA)\cite{wang2024mixture,li2024smoa} allows coordinated reasoning, while SymbolicMoE\cite{chen2025symbolic} aggregates top-k responses based on skill relevance. KnowledgeCard~\cite{feng2024knowledge} dynamically selects smaller fine-tuned agents, and Heterogeneous Swarm~\cite{feng2025heterogeneous} optimizes reasoning via a DAG structure. In contrast to prior work focused on dataset-level heterogeneity, our method targets a more granular challenge: each individual question requires reasoning across multiple subject domains.

\section{Methodology}

\label{sec:method}

\textbf{Method Overview.} As illustrated in Figure~\ref{fig:method}, our proposed S-DAG framework enables subject-aware multi-agent reasoning for complex, multi-subject questions through a two-stage process. In Stage 1, the input question is encoded using a BERT encoder, and a Graph Neural Network predicts relevant subjects and their dependencies to construct a Subject-based Directed Acyclic Graph (S-DAG), filtering out irrelevant domains. In Stage 2, expert LLMs are selected based on subject-specific capability profiles and assigned to the S-DAG nodes. These agents collaborate according to the graph’s structure, with directed edges guiding information flow from supporting to dominant subjects, enabling efficient and accurate multi-subject reasoning.

\subsection{Problem Setup}

Let $\mathcal{Q}$ denotes a natural language question that spans multiple subject areas $\mathcal{S}=\{s_1,s_2,...,s_K\}$, and let $\mathcal{M}=\{M_1,M_2,...,M_n\}$ represent a pool of domain-specific expert LLMs. The objective is to solve $\mathcal{Q}$ by determining a small set of relevant subject domains $\mathcal{S}_Q \subseteq \mathcal{S}$ (typically $|\mathcal{S_Q}|\le 5$), identifying the interdependencies among them, and assigning each subject $s_i \in \mathcal{S_Q}$ to a corresponding expert model $M_j \in \mathcal{M}$ that is most proficient in that domain. 
More specifically, we model the relationships between the selected subjects by constructing a DAG over $\mathcal{S_Q}$, i.e., S-DAG, where a directed edge $s_i \rightarrow s_j$ indicates that subject $s_i$ provides auxiliary support for reasoning in subject $s_j$ in solving $\mathcal{Q}$. This structure reflects the compositional nature of multi-domain reasoning and determines how different experts should collaborate. 
To obtain the S-DAG, we define a fully connected directed graph $\mathcal{G}=\{\mathcal{V},\mathcal{E}\}$, where each node $v_i \in \mathcal{V}$ corresponds to a subject $s_i \in \mathcal{S}$, and each directed edge $(v_i,v_j) \in \mathcal{E}$ encodes a potential dependency between subject pairs. From $\mathcal{G}$, we derive the pruned S-DAG $\mathcal{G_Q}=\{\mathcal{S_Q},\mathcal{A_Q}\}$, which serves as a high-level reasoning blueprint. It guides the selection of a subset of expert LLMs, $\mathcal{M_Q} \subseteq \mathcal{M}$, and defines the collaboration topology among them, enabling effective multi-agent reasoning over complex, interdisciplinary queries.

\subsection{Preprocessing}

\subsubsection{GNN Training.}
\label{sec:GNN-Training}


To effectively solve the multi-subject questions, we employ a set of expert agents, each specializing in a distinct domain. While LLMs are capable of identifying relevant subjects via prompting, they often struggle to capture fine-grained inter-subject dependencies and may produce outputs that are noisy, inconsistent, or lacking in structural coherence. To address this limitation, we introduce a trainable GNN module that learns to model subject dependencies through iterative message-passing over a subject-level graph. The resulting subject graph, or S-DAG, serves as a robust structural prior that guides information flow across agents.

Before training the GNN for S-DAG generation, we preprocess the dataset to construct ground-truth subject graphs for supervision. Given a question $\mathcal{Q}$, we prompt a LLM to extract a set of relevant subject domains $\mathcal{S_Q}=\{s_1,s_2,...,s_k\}$. Each subject $s_i$ is assigned a relevance weight $\{w_i\}_{i=1}^{k} \in [0,1]$, indicating its relative importance for solving $\mathcal{Q}$. Using these subjects and weights, we can construct a ground-truth subject graph $\mathcal{G_Q}=\{\mathcal{S_Q}, \mathcal{A_Q}\}$ for question $\mathcal{Q}$, where $\mathcal{A_Q} \in \{0,1\}^{k\times k}$ is the adjacency matrix. Specifically, a directed edge $a_{\mathcal{Q}}^{ij} = 1$ indicates that subject $s_i$ (with lower weights) supports subject $s_j$ (with higher weight), reflecting the support-to-dominant subject relationship essential for multi-agent subject-specific reasoning. To ensure consistency, we use \texttt{qwen-turbo-0919} ~\cite{yang2024qwen2} as the \textit{Subject LLM} and perform three rounds of processing for each question, only retaining subjects that appear consistently.
Further details on dataset preprocessing and subject graph construction are provided in appendix. 
To ensure comprehensive modeling of potential subject interactions, we define a static, fully connected directed graph $\mathcal{G}=\{\mathcal{V},\mathcal{E}\}$, where each node corresponds to a candidate subject and each edge represents a possible dependency. This graph serves as the structural input for the GNN during S-DAG generation. 


Given a question $\mathcal{Q}$, a pretrained transformer encoder, BERT~\cite{devlin2019bert}, encodes it into a dense vector $\mathbf{h}_\mathcal{Q} \in \mathbb{R}^d$, capturing the semantic intent of the input. Each subject node representation $v \in \mathcal{V}$ is initialized with a fused features of its subject embedding and question embedding via an MLP:

\begin{equation}
    \mathbf{x}_i^{(0)} = \mathrm{MLP}_\text{init}([\mathbf{h}_{i}; \mathbf{h}_{\mathcal{Q}}]).
\end{equation}

The initialized node features are then updated through layers of directional message passing within the GNN. The final node representations are used for joint node and edge prediction, yielding a predicted subject graph $\mathcal{G}_\mathcal{Q}=\{\hat{\mathcal{S}}_\mathcal{Q},\hat{\mathcal{A}}_\mathcal{Q}\}$, where $\hat{s}^i_{\mathcal{Q}} \in \hat{\mathcal{S}}_\mathcal{Q}$ is the predicted relevance score for subject $s_i$, and $\hat{a}_{\mathcal{Q}}^{ij}$ is the predicted edge score for the dependency from $s_i$ to $s_j$. Formally, the model is trained to minimize a multi-task binary cross-entropy loss between the predicted and ground-truth node and edge labels:

\begin{equation}
    \mathcal{L} = \lambda_\text{node}\cdot\sum_{i=1}^{K} \text{BCE}(\hat{s}_\mathcal{Q}^i,s_\mathcal{Q}^i) + \lambda_\text{edge}\cdot \sum_{i\ne j}\text{BCE}(\hat{a}_\mathcal{Q}^{ij},a_\mathcal{Q}^{ij}),
\end{equation}
where $\lambda_\text{node}$ and $\lambda_\text{edge}$ control the weighting between node-level and edge-level supervision. To avoid penalizing irrelevant subjects, edge loss terms are masked when both $s_i=0$ and $s_j=0$. This objective encourages the model to learn subjects and their interaction patterns reflecting the reasoning dependencies in multi-domain tasks. Once trained, the GNN is used at inference time to construct a S-DAG for any new question, guiding the structure and flow of LLM-based agent collaboration tailored to the problem’s subject composition.

\subsubsection{LLMs Subject Capability Profile.}

To optimize multi-agent collaboration, we construct a capability profile for each LLM based on its performance across various subject domains. This profile captures the subject-specific strengths of each model and serves as the foundation for expert selection within the heterogeneous multi-agent system. By leveraging these profiles, we ensure that queries are routed to the most competent models, improving both accuracy and efficiency in multi-domain reasoning.

Unlike prior work~\cite{chen2025symbolic}, which assigns a single subject label to each question, our approach captures the multi-domain nature by assigning weights to all relevant subjects. These weights reflect the relative importance of each subject in solving the question, enabling a more fine-grained and accurate assessment of each LLM's performance across different domains. For instance, if model $M_i$ answers a question spanning math, physics, and biology with weights $\mathrm{\{'math': 0.5, 'physics': 0.3, 'biology': 0.2\}}$ correctly, it would accordingly obtain the performance score $\mathrm{Score}_{M_i} = \mathrm{\{'math': +0.5, 'physics': +0.3, 'biology': +0.2\}}$. By aggregating such weighted scores across a diverse set of questions, each model builds a subject capability profile that accurately reflects its domain expertise.

To construct these profiles, we randomly select 200 questions from the test set to assess LLM performance across each subject domain. The performance scores are then normalized to ensure comparability across domains and models. The normalized capability scores $C_{ij}$ for model $M_i$ and subject $s_j$ are given by:

\begin{equation}
    C_{ij} = \frac{\mathrm{Score}_{M_i,s_j}}{\sum_{s_k \in \mathcal{S}} \mathrm{Score}_{M_i,s_k}},
\end{equation}
where $S$ is the set of all subject domains. Normalization ensures that the sum of scores across all domains for each model equals 1, preventing skewed evaluations and enabling fair comparisons. This profiling system enables dynamic, context-aware model selection based on the subject composition of each query, ensuring that the most suitable expert LLM is invoked during inference.

\subsection{Multi-Agent Collaboration}

\subsubsection{S-DAG Generation.}

During inference, given an input question, we employ the trained GNN model to refine both the question embedding and subject node features. The node and edge classifiers (MLP modules) simultaneously predict relevant subject nodes and their dependencies. This inference procedure mirrors the training phase described in Section~\ref{sec:GNN-Training}. The resulting S-DAG $\mathcal{G_Q}$ is a pruned subgraph of the fully connected subject graph $\mathcal{G}$, retaining only the most relevant subjects and directed relationships. The complete S-DAG generation process is presented in Alg.~\ref{alg:s-dag-gen}.

\textit{Why is S-DAG suitable for guiding multi-agent subject-specific reasoning?}
The S-DAG captures both hierarchical and interdependent relationships among subjects. Dominant subjects represent the central focus of reasoning, while supporting subjects—linked via directed edges—provide complementary knowledge. This structured representation naturally defines a collaboration mechanism among expert agents: support agents supply contextual input that enriches the reasoning of dominant agents. Further discussion of the theoretical motivation and construction principles is provided in appendix.


\begin{algorithm}[t]
\small
\caption{S-DAG Generation} \label{alg:s-dag-gen}
\renewcommand{\algorithmicrequire}{\textbf{Input:}} \renewcommand{\algorithmicensure}{\textbf{Output:}} 

\begin{algorithmic}[1] 

\REQUIRE Input question $\mathcal{Q}$. Randomly initialized node embedding $\{\mathbf{h}_i\}_{i\in \mathcal{V}}$. An initial fully connected graph $\mathcal{G}=\{\mathcal{V},\mathcal{E}\}$.
\ENSURE Generated S-DAG $\mathcal{G_Q}$.

\FOR{each question $\mathcal{Q}$}
    \STATE \textbf{Step 1: Question Embedding}
    \STATE Embed the question: $\mathbf{h}_{\mathcal{Q}} \leftarrow \mathrm{BERT}(\mathcal{Q})$.
    \STATE Initialize node features:
    $\mathbf{x}_{i} \leftarrow \mathrm{MLP}_\text{init}([\mathbf{h}_{i}; \mathbf{h}_{\mathcal{Q}}]), i \in \mathcal{V}$.

    \STATE \textbf{Step 2: S-DAG Generation}
    \STATE Update the node features: $\mathbf{x}_{i} = f_\theta(\mathcal{G},\mathbf{x}_i^{(0)}), i \in \mathcal{V}$.
    \STATE Node prediction:  \( s_{i, i \in \mathcal{V}} \leftarrow \mathrm{MLP_{node}}(\mathbf{x}_i)\).
    \STATE Edge prediction: \( a_{\mathcal{Q}}^{ij, (i,j)\in \mathcal{E}} \leftarrow \mathrm{MLP_{node}}(\mathbf{x}_i;\mathbf{x}_j;\mathbf{h}_{\mathcal{Q}}) \).
    \STATE S-DAG Construction: \( \mathcal{G_Q} = \{s_{\mathcal{Q}}^i,a_{\mathcal{Q}}^{ij}\}_{i=1}^{K} \).
\ENDFOR

\end{algorithmic}
\end{algorithm}

\subsubsection{Multi-Agent Information Flow over S-DAG.}

Given the S-DAG generated in the previous step, the next is to select the appropriate expert LLMs based on their subject proficiency. Based on the LLM subject capability profile process, we match each subject node in the S-DAG to an expert LLM specializing in that domain. If $C_{ij}$ represents the performance score of LLM $M_i$ on subject $s_j$, the LLM selection for a particular subject $s_j$ could be expressed as:

\begin{equation}
    M_j = \arg\max_{M_i} C_{ij}.
\end{equation}

The multi-agent collaboration mechanism is defined by the directed relationships encoded in the S-DAG, as illustrated in Figure \ref{fig:method}. Each edge in the graph represents an information flow dependency between two subject domains, guiding how the associated LLM agents should collaborate. Specifically, if a query involves subjects $s_1$ and $s_2$ with a directed edge from $s_1$ to $s_2$, the output of the agent corresponding to $s_1$ and the original query jointly serve as the prompt input for the agent corresponding to $s_2$, thereby forming a collaborative reasoning pipeline. The prompting strategy that enables this information flow is detailed in appendix. This process can be formalized as:


\begin{equation}
    y_j^{out} = M_j(\{y_i^{out}|a_{ij}=1\}, \mathcal{Q}),
\end{equation}
where $y_i^{out}$ denotes the output of agent $M_i$ associated with subject $s_i$, $a_{ij}=1$ indicates subject $s_i$ has a directed edge to $s_j$ in the S-DAG, and $\mathcal{Q}$ represents the original question, which is included as a shared input to all agents. This dynamic and dependency-driven collaboration enables the system to aggregate reasoning results and progressively refine the final answer through multi-agent cooperation.

\section{Experiments}

\label{sec:exp}

\subsection{Experiment Setup}

\begin{table}[t!]
\small
\centering
\scalebox{0.99}{
\begin{tabular}{lccc}
\toprule
\textbf{Dataset} & \textbf{Train Set} & \textbf{Test Set} & \textbf{Avg. Subject/Q}\\
\midrule
MMLU-Pro & 1173 & 503 & 4.4\\
GPQA & 302 & 129 & 3.8\\
MedMCQA & 396 & 169 & 3.5\\
\bottomrule
\end{tabular}}
\caption{Overview of the curated multi-subject datasets used in our experiments. We manually select questions that span multiple subject areas to better evaluate heterogeneous reasoning capabilities. ``Avg. Subject/Q" denotes the average number of distinct subjects involved per question, reflecting the interdisciplinary complexity of each dataset.}
\label{tab:datasets}
\end{table}

\paragraph{Datasets.} We evaluate our proposed method on three benchmarks. MMLU-Pro~\cite{wang2025mmlu} is a challenging extension of the MMLU benchmark, covering 14 college-level subjects. GPQA~\cite{rein2024gpqa} is a dataset of graduate-level science questions designed to be difficult. MedMCQA~\cite{pal2022medmcqa} is a collection of medical entrance exam questions across 21 subdomains. To better reflect real-world heterogeneous reasoning scenarios, we preprocess each dataset to select samples involve multiple subject areas. We construct a dedicated dataset, as shown in Table \ref{tab:datasets}. We also construct a profiling set with 200 samples used to evaluate the subject-specific capabilities of LLMs. This dataset preprocessing ensures that our evaluation aligns with the core challenge addressed in this paper: selecting and coordinating multiple expert agents to solve complex, multi-subject reasoning problems. The details of the dataset curation are shown in appendix.


\begin{table*}[t!]
\small
\centering
\scalebox{0.95}{
\begin{tabular}{lllcccc}
\toprule
\textbf{Category} & \textbf{Method} & \textbf{Model} & \textbf{MMLU-Pro} & \textbf{GPQA} & \textbf{MedMCQA} & \textbf{Avg.}\\
\midrule
Closed-Source Single Model
& CoT~\cite{wei2022chain} & GPT4o-mini & 49.42~$\pm$~0.27 & 47.31~$\pm$~0.52 & 78.82~$\pm$~0.35 & 58.52 \\
\midrule
Open-Source
& CoT & Qwen2.5 72b & 50.81~$\pm$~0.46 & 48.98~$\pm$~0.35 & 80.44~$\pm$~0.47 & 60.08 \\
Single Model
& CoT & Llama3.3 70b & 51.92~$\pm$~0.39 & 48.83~$\pm$~0.41 & 79.36~$\pm$~0.61 & 60.04 \\
\midrule
\multirow{3}{*}{Single-Model}
& CoT & Qwen2.5 7b & 41.86~$\pm$~0.29 & 44.51~$\pm$~0.58 & 72.07~$\pm$~0.39 & 52.81 \\
& Self-Refine~\cite{madaan2023self} & Qwen2.5 7b & 44.92~$\pm$~0.34 & 43.83~$\pm$~0.12 & 74.58~$\pm$~0.28 & 54.44 \\
& MoE~\cite{zhou2022mixture} & LLM Pool & 42.57~$\pm$~0.55 & 45.67~$\pm$~0.34 & 75.45~$\pm$~0.49 & 54.56 \\
& GraphRouter (Feng et al. 2025) & LLM Pool & 44.94~$\pm$~0.94 & 46.23~$\pm$~0.82 & 76.92~$\pm$~0.72 & 56.03 \\
\midrule
\multirow{3}{*}{Multi-Model}
& MAD~\cite{du2023improving} & Qwen2.5 7b & 45.82~$\pm$~0.13 & \underline{46.81~$\pm$~0.21} & 76.55~$\pm$~0.15 & 56.39 \\
& Symbolic-MoE~\cite{chen2025symbolic} & LLM Pool & \underline{48.13~$\pm$~0.62} & 45.92~$\pm$~0.51 & \textbf{78.55~$\pm$~0.61} & \underline{57.53} \\
& {\bf S-DAG (Ours)} & LLM Pool & \textbf{50.98~$\pm$~0.19} & \textbf{49.82~$\pm$~0.24} & \underline{78.38~$\pm$~0.35} & \textbf{59.73} \\
\bottomrule
\end{tabular}}
\caption{Performance comparison of single-model and multi-model approaches on MMLU-Pro, GPQA, and MedMCQA. We compare various baselines across closed-source, open-source, single-agent and multi-agent settings. We \textbf{bold} the best results and \underline{underline} the second-best (excluding methods using bigger or proprietary models).}
\label{tab:acc}
\end{table*}

\paragraph{LLM Pool with Various Experts.} To enable subject-aware reasoning and fine-grained agent specialization, we construct a pool of domain-specific expert LLMs. Each model is either pretrained or fine-tuned on data aligned with a specific academic or professional domain, such as mathematics, medicine, law, or economics. These models, typically ranging from 7B to 13B parameters, are computationally efficient and well-suited for multi-agent composition. In total, we select 14 expert LLMs spanning a broad range of disciplines. Appendix provides details on the expert models, corresponding subject domains, and Hugging Face links. For instance, \texttt{DeepseekMath} is used for mathematics, while \texttt{BioMistral} is assigned to biology. These expert LLMs form the foundation of our multi-agent system, where each agent is instantiated from the most suitable LLM based on the subject assignments derived from the S-DAG and LLM identification process.


\paragraph{Baselines.} Our selection of baselines is guided by the goal of evaluating the challenge of heterogeneous reasoning, where a single complex question spans multiple subject domains. We consider two primary categories. First, we evaluate single-model methods to test whether a single, high-performance, general-purpose LLM can effectively handle multi-subject reasoning. Specifically, we include models such as closed-source GPT4o-mini~\cite{hurst2024gpt}, open-source Qwen2.5-72B~\cite{yang2024qwen2} and Llama3.3-70B~\cite{grattafiori2024llama}. Then, the Mixture-of-Experts (MoE) approaches that dynamically select one expert model; GraphRouter that utilize a GNN to select expert model based on contextual information given a problem. Second, we explore multi-model methods, which leverage specialized expertise through collaborative or modular strategies. These include Multi-Agent Debate (MAD), where multiple agents reason through dialogue~\cite{liang2024encouraging};  and Symbolic-MoE~\cite{chen2025symbolic}, which conducts skill-level expert selection. These baselines provide a comprehensive framework to assess both generalist and specialist strategies for tackling complex, interdisciplinary reasoning tasks.

\paragraph{Implementation Details.} All experiments, including baselines and our proposed method, are conducted using A100 GPUs with 40 GB memory. For large models such as the 70B open-source LLMs and Qwen variants, we use API-based inference, while smaller expert models in the LLM pool are deployed locally. The decoding temperature is set to 0.7, and the maximum output length is fixed at 4096 tokens across all LLMs. Adam optimizer is used to train GNN and MLP models, and the seed is fixed. Results are averaged across three trials, and we compute the standard deviations as the statistical indicator. 
Details on model selection and the LLM pool are provided in appendix.


\subsection{Results}

The main empirical findings are shown in Table~\ref{tab:acc}. We evaluate the performance of various reasoning frameworks across selected samples from MMLU-Pro, GPQA, and MedMCQA, including both single-model and multi-model settings, as well as open-source and closed-source configurations. The average accuracy is acquired by averaging results across the three benchmarks. Below, we highlight key insights derived from this comparative analysis.

\paragraph{Superior Accuracy of S-DAG.}
Our S-DAG framework achieves the highest average accuracy (59.73\%). It consistently outperforms both single-model and multi-model baselines. In particular, compared to single-model approaches that rely on expert selection—such as MoE (54.56\%) and GraphRouter (56.03\%)—S-DAG demonstrates a significant improvement. This result underscores the benefit of explicitly modeling inter-subject dependencies, rather than relying solely on selection-based mechanisms.


\paragraph{Robustness over Multi-Model Baselines.}
Our S-DAG also outperforms several competitive multi-agent systems. It exceeds the performance of Symbolic-MoE (57.53\%) and MAD (56.39\%) in average accuracy. Although Symbolic-MoE shows strong performance on MedMCQA (78.55\%), its effectiveness does not consistently translate across other benchmarks. In contrast, S-DAG maintains a well-balanced performance across all three tasks, demonstrating its robustness and adaptability in coordinating specialized agents across diverse reasoning scenarios.


\begin{table*}[t!]
\small
\centering
\scalebox{0.95}{
\begin{tabular}{lcccccc}
\toprule
\textbf{Variant} & \textbf{GNN Coord.} & \textbf{Model Selection} & \textbf{Graph Structure} & \textbf{Avg. Accuracy (\%)} & \textbf{Inf. Time (s)} & \textbf{\# LLM Calls} \\
\midrule

w/o GNN, random model & \XSolidBrush & \XSolidBrush & S-DAG & 41.12 & \textbf{14.21} & 5.1\\
w/ GNN, random model & \Checkmark & \XSolidBrush & S-DAG & 42.19 & 14.82 & 4.1\\
w/o GNN, profiled model & \XSolidBrush & \Checkmark & S-DAG & 53.51 & 14.53 & 5.1\\
Fully Connected Graph & \Checkmark & \Checkmark & Fully-connected & 57.29 & 38.45 & 8.2\\
\textbf{S-DAG (Ours)} & \Checkmark & \Checkmark & S-DAG & \textbf{59.73} & 15.02 & \textbf{4.1}\\
\bottomrule
\end{tabular}}
\caption{Ablation study on the effects of coordination (GNN), model selection (LLM profiling), and graph structure. Accuracy is averaged across MMLU-Pro, GPQA, and MedMCQA. Inference efficiency is measured via average inference time and the number of LLM calls per instance.}
\label{tab:ablation}
\end{table*}

\paragraph{Competitiveness with Large LLMs.}
Despite being composed of smaller, domain-specific expert models, S-DAG achieves performance competitive with large-scale, monolithic LLMs. It surpasses the closed-source GPT-4o-mini (58.52\%) and closely matches the performance of open-source leaders such as Qwen2.5 72B (60.08\%) and Llama3.3 70B (60.04\%). This indicates that structured coordination among lightweight experts can rival or exceed the capabilities of significantly larger models, offering a more efficient and cost-effective solution for complex reasoning tasks.

\subsection{Ablation Study}

To quantify the contribution of individual components in our S-DAG framework, we conduct a structured ablation study, as presented in Table~\ref{tab:ablation}. This analysis isolates three key design factors: (1) the presence or absence of GNN-based coordination, (2) the use of subject-aware model selection via LLM profiling versus random assignment, and (3) the impact of graph topology, comparing our sparse S-DAG structure to a fully connected alternative. We evaluate each configuration across two core dimensions: task performance and computational efficiency. Inference time denotes the average wall-clock latency required to process a single multi-subject question, encompassing decomposition, agent invocation, message passing, and response synthesis. LLM call count indicates the average number of distinct language model calls per instance, serving as a proxy for computational and monetary cost.


\begin{figure}[t!]
    \small
    \centering
    \includegraphics[width=0.96\linewidth]{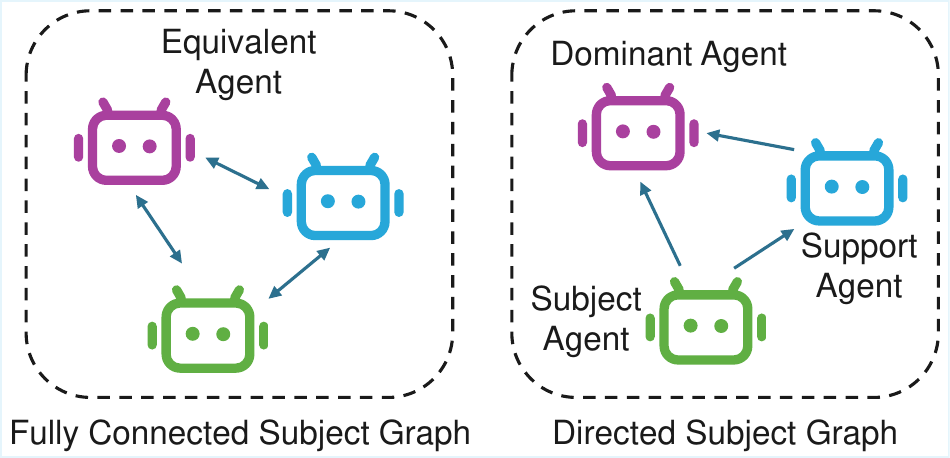}
    \caption{Comparison between the Fully Connected Graph and our S-DAG.}
    \label{fig:directed}
\end{figure}

\paragraph{Effectiveness of GNN Module.} Rather than directly prompting the LLM and construct the S-DAG manually, we introduce a GNN module that learns subject interdependencies from data and automatically produces the S-DAG. To evaluate the effectiveness of this design choice, we conduct an ablation study comparing our learned S-DAG with one derived purely from LLM-generated subject weights. The results show that using the LLM-generated S-DAG yields a significantly lower accuracy compared to 59.73\% achieved by our GNN-based approach. This performance gap highlights the limitations of relying solely on LLM outputs, which can be noisy and inconsistent. In contrast, our GNN module leverages training data to learn robust and context-sensitive subject relationships, resulting in more accurate and concise graph structures for multi-agent reasoning.

\paragraph{Effectiveness of LLM Profile.} To assess the benefit of our LLM profiling strategy, we compare our subject-aware model selection approach against a baseline that randomly selects expert models for each subject node in the S-DAG. The results demonstrate that using our profiled subject–model matching significantly improves performance across all benchmarks. This highlights the importance of aligning subject-specific tasks with LLMs that exhibit strong domain expertise. Without profiling, the system risks assigning questions to suboptimal models, which can lead to degraded reasoning quality and inconsistent outputs.

\paragraph{Effectiveness of Directed Acyclic Graph Structure.} To evaluate the impact of graph topology, we compare our S-DAG with a fully connected graph (FCG) variant. As illustrated in Figure~\ref{fig:directed}, the FCG allows unrestricted bidirectional communication among all subject nodes, which leads to over-communication and redundant information flow. In contrast, our S-DAG enforces a sparse, hierarchical structure that streamlines reasoning. Empirically, S-DAG achieves higher accuracy (59.73\%) while significantly reducing inference time (15.02s) and the number of LLM calls (4.1 per instance). These results demonstrate that fully connected communication is suboptimal for multi-agent reasoning, and that structured, directional coordination leads to more efficient and effective performance.

\section{Conclusion}

\label{sec:conclusion}

We present S-DAG, a novel framework for heterogeneous reasoning that leverages fine-grained subject-level analysis to guide multi-agent collaboration. By constructing a subject-based directed acyclic graph via a GNN, our method captures subject interdependencies, enabling targeted coordination among specialized LLMs. This structured approach yields efficient, subject-level reasoning. Experiments on three challenging benchmarks show that S-DAG consistently outperforms both single-model and multi-model baselines, achieving competitive accuracy with large-scale LLMs at significantly lower computational cost. Ablation studies further highlight the benefits of the DAG structure in enhancing performance and efficiency.

\section{Acknowledgments}
This research was supported by Project P0049179 under the Innovation and Technology Fund – Guangdong–Hong Kong Technology Cooperation Funding Scheme (ITF-TCFS), funded by the Innovation and Technology Commission (Funding Body Ref. No. GHP/386/23SZ).

\bibliography{aaai2026}

@inproceedings{feng2025graphrouter,
title = {GraphRouter: A Graph-based Router for {LLM} Selections},
author={Tao Feng and Yanzhen Shen and Jiaxuan You},
booktitle = {The Thirteenth International Conference on Learning Representations},
year={2025},
url={https://openreview.net/forum?id=eU39PDsZtT}
}

@article{chen2025symbolic,
  title = {Symbolic Mixture-of-Experts: Adaptive Skill-based Routing for Heterogeneous Reasoning},
  author={Chen, Justin Chih-Yao and Yun, Sukwon and Stengel-Eskin, Elias and Chen, Tianlong and Bansal, Mohit},
  journal = {arXiv preprint arXiv:2503.05641},
  year={2025}
}

@inproceedings{
wu2024autogen,
title = {AutoGen: Enabling Next-Gen {LLM} Applications via Multi-Agent Conversations},
author={Qingyun Wu and Gagan Bansal and Jieyu Zhang and Yiran Wu and Beibin Li and Erkang Zhu and Li Jiang and Xiaoyun Zhang and Shaokun Zhang and Jiale Liu and Ahmed Hassan Awadallah and Ryen W White and Doug Burger and Chi Wang},
booktitle = {First Conference on Language Modeling},
year={2024},
url={https://openreview.net/forum?id=BAakY1hNKS}
}

@article{yue2025survey,
  title = {A Survey of Large Language Model Agents for Question Answering},
  author={Yue, Murong},
  journal = {arXiv preprint arXiv:2503.19213},
  year={2025}
}

@article{zhuang2023toolqa,
  title = {Toolqa: A Dataset for Llm Question Answering With External Tools},
  author={Zhuang, Yuchen and Yu, Yue and Wang, Kuan and Sun, Haotian and Zhang, Chao},
  journal = {Advances in Neural Information Processing Systems},
  volume={36},
  pages={50117--50143},
  year={2023}
}

@article{huang2023grounded,
  title = {Grounded Decoding: Guiding Text Generation With Grounded Models for Embodied Agents},
  author={Huang, Wenlong and Xia, Fei and Shah, Dhruv and Driess, Danny and Zeng, Andy and Lu, Yao and Florence, Pete and Mordatch, Igor and Levine, Sergey and Hausman, Karol and others},
  journal = {Advances in Neural Information Processing Systems},
  volume={36},
  pages={59636--59661},
  year={2023}
}

@article{ke2025survey,
  title = {A Survey of Frontiers in LLM Reasoning: Inference Scaling, Learning To Reason, And Agentic Systems},
  author={Ke, Zixuan and Jiao, Fangkai and Ming, Yifei and Nguyen, Xuan-Phi and Xu, Austin and Long, Do Xuan and Li, Minzhi and Qin, Chengwei and Wang, Peifeng and Savarese, Silvio and others},
  journal = {arXiv preprint arXiv:2504.09037},
  year={2025}
}

@article{zhang2024llm,
  title = {Llm As A Mastermind: A Survey of Strategic Reasoning With Large Language Models},
  author={Zhang, Yadong and Mao, Shaoguang and Ge, Tao and Wang, Xun and de Wynter, Adrian and Xia, Yan and Wu, Wenshan and Song, Ting and Lan, Man and Wei, Furu},
  journal = {arXiv preprint arXiv:2404.01230},
  year={2024}
}

@article{guo2024large,
  title = {Large Language Model Based Multi-agents: A Survey of Progress And Challenges},
  author={Guo, Taicheng and Chen, Xiuying and Wang, Yaqi and Chang, Ruidi and Pei, Shichao and Chawla, Nitesh V and Wiest, Olaf and Zhang, Xiangliang},
  journal = {arXiv preprint arXiv:2402.01680},
  year={2024}
}

@inproceedings{du2023improving,
  title = {Improving Factuality And Reasoning in Language Models Through Multiagent Debate},
  author={Du, Yilun and Li, Shuang and Torralba, Antonio and Tenenbaum, Joshua B and Mordatch, Igor},
  booktitle = {Forty-first International Conference on Machine Learning},
  year={2023}
}

@article{mumuni2025large,
  title = {Large Language Models for Artificial General Intelligence (AGI): A Survey of Foundational Principles And Approaches},
  author={Mumuni, Alhassan and Mumuni, Fuseini},
  journal = {arXiv preprint arXiv:2501.03151},
  year={2025}
}

@article{kojima2022large,
  title = {Large Language Models Are Zero-shot Reasoners},
  author={Kojima, Takeshi and Gu, Shixiang Shane and Reid, Machel and Matsuo, Yutaka and Iwasawa, Yusuke},
  journal = {Advances in Neural Information Processing Systems},
  volume={35},
  pages={22199--22213},
  year={2022}
}

@article{feng2025one,
  title = {When One LLM Drools, Multi-LLM Collaboration Rules},
  author={Feng, Shangbin and Ding, Wenxuan and Liu, Alisa and Wang, Zifeng and Shi, Weijia and Wang, Yike and Shen, Zejiang and Han, Xiaochuang and Lang, Hunter and Lee, Chen-Yu and others},
  journal = {arXiv preprint arXiv:2502.04506},
  year={2025}
}

@inproceedings{hoffmann2022training,
  title = {Training Compute-optimal Large Language Models},
  author={Hoffmann, Jordan and Borgeaud, Sebastian and Mensch, Arthur and Buchatskaya, Elena and Cai, Trevor and Rutherford, Eliza and de Las Casas, Diego and Hendricks, Lisa Anne and Welbl, Johannes and Clark, Aidan and others},
  booktitle = {Proceedings of the 36th International Conference on Neural Information Processing Systems},
  pages={30016--30030},
  year={2022}
}

@article{talebirad2023multi,
  title = {Multi-agent Collaboration: Harnessing the Power of Intelligent Llm Agents},
  author={Talebirad, Yashar and Nadiri, Amirhossein},
  journal = {arXiv preprint arXiv:2306.03314},
  year={2023}
}

@article{han2024llm,
  title = {LLM Multi-agent Systems: Challenges And Open Problems},
  author={Han, Shanshan and Zhang, Qifan and Yao, Yuhang and Jin, Weizhao and Xu, Zhaozhuo and He, Chaoyang},
  journal = {arXiv preprint arXiv:2402.03578},
  year={2024}
}

@article{masoudnia2014mixture,
  title = {Mixture of Experts: A Literature Survey},
  author={Masoudnia, Saeed and Ebrahimpour, Reza},
  journal = {Artificial Intelligence Review},
  volume={42},
  pages={275--293},
  year={2014},
  publisher={Springer}
}

@article{zhou2022mixture,
  title = {Mixture-of-experts With Expert Choice Routing},
  author={Zhou, Yanqi and Lei, Tao and Liu, Hanxiao and Du, Nan and Huang, Yanping and Zhao, Vincent and Dai, Andrew M and Le, Quoc V and Laudon, James and others},
  journal = {Advances in Neural Information Processing Systems},
  volume={35},
  pages={7103--7114},
  year={2022}
}

@article{cai2024survey,
  title = {A Survey on Mixture of Experts},
  author={Cai, Weilin and Jiang, Juyong and Wang, Fan and Tang, Jing and Kim, Sunghun and Huang, Jiayi},
  journal = {arXiv preprint arXiv:2407.06204},
  year={2024}
}

@article{wang2024mixture,
  title = {Mixture-of-agents Enhances Large Language Model Capabilities},
  author={Wang, Junlin and Wang, Jue and Athiwaratkun, Ben and Zhang, Ce and Zou, James},
  journal = {arXiv preprint arXiv:2406.04692},
  year={2024}
}

@article{li2024smoa,
  title = {Smoa: Improving Multi-agent Large Language Models With Sparse Mixture-of-agents},
  author={Li, Dawei and Tan, Zhen and Qian, Peijia and Li, Yifan and Chaudhary, Kumar Satvik and Hu, Lijie and Shen, Jiayi},
  journal = {arXiv preprint arXiv:2411.03284},
  year={2024}
}

@article{zhang2024chain,
  title = {Chain of Agents: Large Language Models Collaborating on Long-context Tasks},
  author={Zhang, Yusen and Sun, Ruoxi and Chen, Yanfei and Pfister, Tomas and Zhang, Rui and Arik, Sercan},
  journal = {Advances in Neural Information Processing Systems},
  volume={37},
  pages={132208--132237},
  year={2024}
}

@inproceedings{liang2024encouraging,
  title = {Encouraging Divergent Thinking in Large Language Models Through Multi-Agent Debate},
  author={Liang, Tian and He, Zhiwei and Jiao, Wenxiang and Wang, Xing and Wang, Yan and Wang, Rui and Yang, Yujiu and Shi, Shuming and Tu, Zhaopeng},
  booktitle = {Proceedings of the 2024 Conference on Empirical Methods in Natural Language Processing},
  pages={17889--17904},
  year={2024}
}

@inproceedings{yin2023exchange,
  title = {Exchange-of-Thought: Enhancing Large Language Model Capabilities Through Cross-Model Communication},
  author={Yin, Zhangyue and Sun, Qiushi and Chang, Cheng and Guo, Qipeng and Dai, Junqi and Huang, Xuan-Jing and Qiu, Xipeng},
  booktitle = {Proceedings of the 2023 Conference on Empirical Methods in Natural Language Processing},
  pages={15135--15153},
  year={2023}
}

@inproceedings{gu2025explain,
  title = {Explain-Analyze-Generate: A Sequential Multi-Agent Collaboration Method for Complex Reasoning},
  author={Gu, WenYuan and Han, Jiale and Wang, HaoWen and Li, Xiang and Cheng, Bo},
  booktitle = {Proceedings of the 31st International Conference on Computational Linguistics},
  pages={7127--7140},
  year={2025}
}

@inproceedings{li2024improving,
  title = {Improving Multi-Agent Debate With Sparse Communication Topology},
  author={Li, Yunxuan and Du, Yibing and Zhang, Jiageng and Hou, Le and Grabowski, Peter and Li, Yeqing and Ie, Eugene},
  booktitle = {Findings of the Association for Computational Linguistics: EMNLP 2024},
  pages={7281--7294},
  year={2024}
}

@inproceedings{tao2025chain,
  title = {Chain-of-Discussion: A Multi-Model Framework for Complex Evidence-Based Question Answering},
  author={Tao, Mingxu and Zhao, Dongyan and Feng, Yansong},
  booktitle = {Proceedings of the 31st International Conference on Computational Linguistics},
  pages={11070--11085},
  year={2025}
}

@inproceedings{zhuge2024gptswarm,
  title = {Gptswarm: Language Agents As Optimizable Graphs},
  author={Zhuge, Mingchen and Wang, Wenyi and Kirsch, Louis and Faccio, Francesco and Khizbullin, Dmitrii and Schmidhuber, J{\"u}rgen},
  booktitle = {Forty-first International Conference on Machine Learning},
  year={2024}
}

@article{feng2025heterogeneous,
  title = {Heterogeneous Swarms: Jointly Optimizing Model Roles And Weights for Multi-LLM Systems},
  author={Feng, Shangbin and Wang, Zifeng and Goyal, Palash and Wang, Yike and Shi, Weijia and Xia, Huang and Palangi, Hamid and Zettlemoyer, Luke and Tsvetkov, Yulia and Lee, Chen-Yu and others},
  journal = {arXiv preprint arXiv:2502.04510},
  year={2025}
}

@article{zhang2024cut,
  title = {Cut the Crap: An Economical Communication Pipeline for Llm-based Multi-agent Systems},
  author={Zhang, Guibin and Yue, Yanwei and Li, Zhixun and Yun, Sukwon and Wan, Guancheng and Wang, Kun and Cheng, Dawei and Yu, Jeffrey Xu and Chen, Tianlong},
  journal = {arXiv preprint arXiv:2410.02506},
  year={2024}
}

@inproceedings{liu2024dynamic,
  title = {A Dynamic LLM-powered Agent Network for Task-oriented Agent Collaboration},
  author={Liu, Zijun and Zhang, Yanzhe and Li, Peng and Liu, Yang and Yang, Diyi},
  booktitle = {First Conference on Language Modeling},
  year={2024}
}

@inproceedings{
feng2024knowledge,
title = {Knowledge Card: Filling {LLM}s' Knowledge Gaps with Plug-in Specialized Language Models},
author={Shangbin Feng and Weijia Shi and Yuyang Bai and Vidhisha Balachandran and Tianxing He and Yulia Tsvetkov},
booktitle = {The Twelfth International Conference on Learning Representations},
year={2024},
url={https://openreview.net/forum?id=WbWtOYIzIK}
}

@article{xin2024atomr,
  title = {AtomR: Atomic Operator-Empowered Large Language Models for Heterogeneous Knowledge Reasoning},
  author={Xin, Amy and Liu, Jinxin and Yao, Zijun and Lee, Zhicheng and Cao, Shulin and Hou, Lei and Li, Juanzi},
  journal = {arXiv preprint arXiv:2411.16495},
  year={2024}
}

@article{
chen2024frugalgpt,
title = {Frugal{GPT}: How to Use Large Language Models While Reducing Cost and Improving Performance},
author={Lingjiao Chen and Matei Zaharia and James Zou},
journal = {Transactions on Machine Learning Research},
issn={2835-8856},
year={2024},
url={https://openreview.net/forum?id=cSimKw5p6R},
note={}
}

@article{yue2025masrouter,
  title = {Masrouter: Learning To Route Llms for Multi-agent Systems},
  author={Yue, Yanwei and Zhang, Guibin and Liu, Boyang and Wan, Guancheng and Wang, Kun and Cheng, Dawei and Qi, Yiyan},
  journal = {arXiv preprint arXiv:2502.11133},
  year={2025}
}

@article{yang2024qwen2,
  title = {Qwen2. 5 Technical Report},
  author={Yang, An and Yang, Baosong and Zhang, Beichen and Hui, Binyuan and Zheng, Bo and Yu, Bowen and Li, Chengyuan and Liu, Dayiheng and Huang, Fei and Wei, Haoran and others},
  journal = {arXiv preprint arXiv:2412.15115},
  year={2024}
}

@article{wang2025mmlu,
  title = {MMLU-Pro: A More Robust And Challenging Multi-Task Language Understanding Benchmark},
  author={Wang, Yubo and Ma, Xueguang and Zhang, Ge and Ni, Yuansheng and Chandra, Abhranil and Guo, Shiguang and Ren, Weiming and Arulraj, Aaran and He, Xuan and Jiang, Ziyan and others},
  journal = {Advances in Neural Information Processing Systems},
  volume={37},
  pages={95266--95290},
  year={2025}
}

@inproceedings{rein2024gpqa,
  title = {Gpqa: A Graduate-level Google-proof Q\&a Benchmark},
  author={Rein, David and Hou, Betty Li and Stickland, Asa Cooper and Petty, Jackson and Pang, Richard Yuanzhe and Dirani, Julien and Michael, Julian and Bowman, Samuel R},
  booktitle = {First Conference on Language Modeling},
  year={2024}
}

@inproceedings{pal2022medmcqa,
  title = {Medmcqa: A Large-scale Multi-subject Multi-choice Dataset for Medical Domain Question Answering},
  author={Pal, Ankit and Umapathi, Logesh Kumar and Sankarasubbu, Malaikannan},
  booktitle = {Conference on Health, Inference, And Learning},
  pages={248--260},
  year={2022},
  organization={PMLR}
}

@article{grattafiori2024llama,
  title = {The Llama 3 Herd of Models},
  author={Grattafiori, Aaron and Dubey, Abhimanyu and Jauhri, Abhinav and Pandey, Abhinav and Kadian, Abhishek and Al-Dahle, Ahmad and Letman, Aiesha and Mathur, Akhil and Schelten, Alan and Vaughan, Alex and others},
  journal = {arXiv preprint arXiv:2407.21783},
  year={2024}
}

@article{hurst2024gpt,
  title = {Gpt-4o System Card},
  author={Hurst, Aaron and Lerer, Adam and Goucher, Adam P and Perelman, Adam and Ramesh, Aditya and Clark, Aidan and Ostrow, AJ and Welihinda, Akila and Hayes, Alan and Radford, Alec and others},
  journal = {arXiv preprint arXiv:2410.21276},
  year={2024}
}

@article{madaan2023self,
  title = {Self-refine: Iterative Refinement With Self-feedback},
  author={Madaan, Aman and Tandon, Niket and Gupta, Prakhar and Hallinan, Skyler and Gao, Luyu and Wiegreffe, Sarah and Alon, Uri and Dziri, Nouha and Prabhumoye, Shrimai and Yang, Yiming and others},
  journal = {Advances in Neural Information Processing Systems},
  volume={36},
  pages={46534--46594},
  year={2023}
}

@article{wei2022chain,
  title = {Chain-of-thought Prompting Elicits Reasoning in Large Language Models},
  author={Wei, Jason and Wang, Xuezhi and Schuurmans, Dale and Bosma, Maarten and Xia, Fei and Chi, Ed and Le, Quoc V and Zhou, Denny and others},
  journal = {Advances in Neural Information Processing Systems},
  volume={35},
  pages={24824--24837},
  year={2022}
}

@inproceedings{devlin2019bert,
  title = {Bert: Pre-training of Deep Bidirectional Transformers for Language Understanding},
  author={Devlin, Jacob and Chang, Ming-Wei and Lee, Kenton and Toutanova, Kristina},
  booktitle = {Proceedings of the 2019 Conference of the North American Chapter of the Association for Computational Linguistics: Human Language Technologies, Volume 1 (long And Short Papers)},
  pages={4171--4186},
  year={2019}
}

@inproceedings{li2025graphs,
title={Graphs Help Graphs: Multi-Agent Graph Socialized Learning},
author={Jialu Li and Yu Wang and Pengfei Zhu and Wanyu Lin and Xinjie Yao and Qinghua Hu},
booktitle={The Thirty-ninth Annual Conference on Neural Information Processing Systems},
year={2025},
url={https://openreview.net/forum?id=lkw2WJLdbh}
}

@inproceedings{yao2025socialized,
title={Socialized Coevolution: Advancing a Better World through Cross-Task Collaboration},
author={Xinjie Yao and Yu Wang and Pengfei Zhu and Wanyu Lin and Ruipu Zhao and Zhoupeng Guo and Weihao Li and Qinghua Hu},
booktitle={Forty-second International Conference on Machine Learning},
year={2025},
url={https://openreview.net/forum?id=0WQJ6DFSKp}
}

@article{lin2025graph,
  title={Graph-Relational Federated Learning: Enhanced Personalization and Robustness},
  author={Lin, Wanyu and Lan, Hao and He, Hao and Li, Baochun},
  journal={IEEE Transactions on Dependable and Secure Computing},
  year={2025},
  publisher={IEEE}
}

\appendix
\onecolumn


\setcounter{secnumdepth}{2}

\section{Multi-Subject Dataset Curation}
\label{app:dataset-preprocessing}

\subsection{Question Analysis}

While benchmark datasets like MMLU-Pro offer broad subject coverage, individual questions within these datasets often pertain to a single domain, limiting their utility for evaluating complex, interdisciplinary reasoning. In contrast, our work targets multi-subject reasoning, where answering a question requires synthesizing knowledge across multiple subject areas. This distinction is critical for advancing generalist models that more closely resemble human cognitive abilities.

To construct a dataset that captures this complexity, we introduce a rigorous preprocessing pipeline aimed at filtering and enhancing questions that truly require interdisciplinary reasoning. For each dataset, we employ the large language model \texttt{qwen-turbo-0919}~\cite{yang2024qwen2} to analyze the subject composition of every question. The model identifies relevant subject domains and assigns a relevance weight to each, reflecting its importance to the question. To ensure robustness and mitigate the variability of model outputs, each question is analyzed three independent times. We then retain only the subjects that consistently appear in all three runs, thereby filtering out spurious or weak associations. This consensus-based filtering increases the precision of subject attribution. The resulting subject weights are normalized across the retained set to maintain comparability and interpretability.

\begin{table*}[htbp]
\small
\caption{Prompt for dataset preprocessing. LLM analyzes the question and outputs the relevant subjects and their weights.}

  \label{tab:question-analysis}
  \centering
  
    \scalebox{0.99}{
  \begin{tabular}{p{13cm}}
    \toprule
    \textbf{Prompt} \\
    \midrule
    Question: \{Q\}
    What are the core knowledge, subjects or skills needed to solve this problem? List 2-5 keywords separated in comma, with the weights (0$\sim$1.0). These weights represent the proportion of these skills are needed in the question. And the proportion of all keywords sum to 1. Candidate keywords: Math, Physics, Chemistry, Law, Engineering, Economics, Health, Psychology, Business, Biology, Philosophy, Computer Science, History, Medicine, Other. Give ONLY the keywords with weights, no other words or explanation.
    Please follow this format: Keywords: $\mathrm{<Math 0.6>, <Physics 0.3>, <Chemistry 0.1>...}$ \\

    \bottomrule
    \end{tabular}}
  
\end{table*}

Overall, this subject attribution framework allows us to curate a high-quality dataset that better reflects real-world, cross-disciplinary reasoning tasks. The prompt template used for subject decomposition and analysis is detailed in Table~\ref{tab:question-analysis}.

We partition the curated dataset into a training set and a test set for model development and evaluation, respectively. Additionally, we reserve a separate set of 200 samples as the profiling set, which is specifically used to assess the subject-specific capabilities of each LLM. This profiling process enables accurate subject–model matching during multi-agent inference.

\begin{figure}[htbp]
    \small
    \centering
    \includegraphics[width=0.5\linewidth]{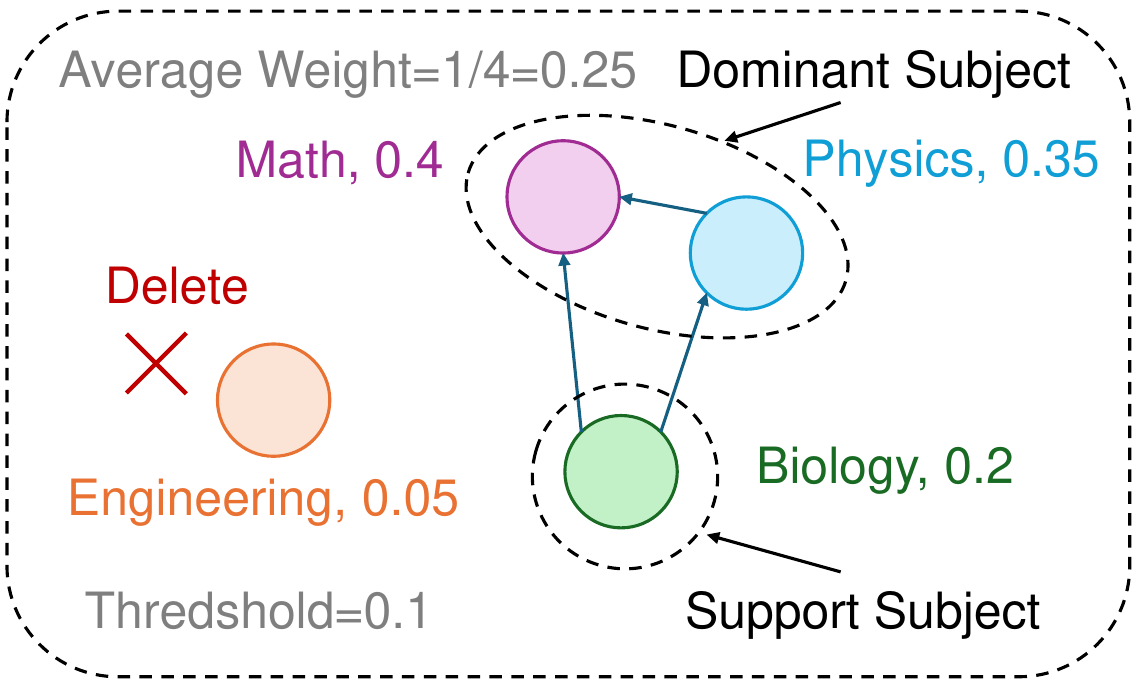}
    \caption{The DAG construction given subjects and their corresponding weights.}
    \label{fig:DAG-construction}
\end{figure}

\subsection{S-DAG Construction}
\label{app:S-DAG Construction based on Problem Analysis}

Once a question is analyzed, we construct a Subject Dependency Acyclic Graph (S-DAG) to represent its ground-truth label. How is the S-DAG constructed based on the subjects and their associated weights? A single question may involve multiple subjects, although it typically centers on a dominant domain. For example, a complex chemistry problem may also require mathematical and biological knowledge. In such cases, chemistry is considered the dominant subject, while mathematics and biology serve as supporting subjects.

To model the flow of knowledge among these subjects, we define directed edges in the S-DAG from supporting subjects to the dominant subject. This directionality reflects the information flow from subjects with lower weights (indicating less emphasis) to those with higher weights (indicating greater emphasis). Consequently, the S-DAG captures the hierarchical and integrative structure of subject dependencies within the question.

In practice, subjects with weights below a predefined threshold (e.g., 0.1) are considered negligible and are discarded. Among the remaining subjects, those with weights exceeding the average (e.g., $1/4 = 0.25$ for four subjects) are designated as dominant subjects. The rest are treated as supporting subjects and are connected to the dominant nodes via directed edges, reflecting their auxiliary role in the reasoning process. An example of S-DAG construction is illustrated in Figure~\ref{fig:DAG-construction}.

\begin{algorithm}[t!]
\small
\caption{GNN Training for S-DAG Generation} \label{alg:gnn_training}
\renewcommand{\algorithmicrequire}{\textbf{Input:}} \renewcommand{\algorithmicensure}{\textbf{Output:}} 

\begin{algorithmic}[1]

\REQUIRE Training dataset $\mathcal{D} = \{(\mathcal{Q}_i, \mathcal{G}_\mathcal{Q}^i)\}_{i=1}^N$, where $\mathcal{Q}_i$ is a question and $\mathcal{G}_\mathcal{Q}^i=\{\mathcal{S}_\mathcal{Q}^i,\mathcal{A}_\mathcal{Q}^i\}$ is the ground truth S-DAG. Randomly initialized node embedding $\{\mathbf{h}_i\}_{i\in \mathcal{V}}$. An initial fully connected graph $\mathcal{G}=\{\mathcal{V},\mathcal{E}\}$.
\ENSURE Trained GNN generator module $f_\theta$

\FOR{each sample $(\mathcal{Q},\mathcal{G}_\mathcal{Q})\in \mathcal{D}$}
    \STATE \textbf{Step 1: Question Embedding}
    \STATE Embed the question: $\mathbf{h}_{\mathcal{Q}} \leftarrow \mathrm{BERT}(\mathcal{Q})$
    \STATE Initialize features of subject nodes: $\mathbf{x}_i^{(0)} \leftarrow \mathrm{MLP}_\text{init}([\mathbf{h}_{i}; \mathbf{h}_{\mathcal{Q}}])$ for $i \in \mathcal{V}$

    \STATE \textbf{Step 2: S-DAG Generation}
    
    \STATE Update the node features: $\mathbf{x}_i = f_\theta(\mathcal{G},\mathbf{x}_i^{(0)})$
    
    \STATE Node prediction:  \( \hat{s}_{\mathcal{Q}}^i \leftarrow \mathrm{MLP_{node}}(\mathbf{x}_i) \) for each \( i \in \mathcal{V} \)

    \STATE Edge prediction: \( \hat{a}_{\mathcal{Q}}^{ij} \leftarrow \mathrm{MLP_{node}}(\mathbf{x}_i;\mathbf{x}_j;\mathbf{h}_{\mathcal{Q}_i}) \) for each \( (i, j) \in \mathcal{E} \)

    \STATE \textbf{Step 3: Compute Loss and Gradient Update}
    \STATE Loss for node and edge prediction: 
    $\mathcal{L} = \lambda_\text{node} \cdot \text{BCE}(\hat{s}_{\mathcal{Q}}^i,s_{\mathcal{Q}}^i) + \lambda_\text{edge} \cdot\text{BCE}(\hat{a}_{\mathcal{Q}}^{ij},a_{\mathcal{Q}}^{ij})$ for each \( i \in \mathcal{V} \) and each \( (i, j) \in \mathcal{E} \)
    \STATE Backpropagate the loss $\mathcal{L}$ and update the model parameters \( f_\theta, \mathrm{MLP_{init}}, \mathrm{MLP_{node}, \mathrm{MLP_{edge}}} \)
\ENDFOR

\end{algorithmic}
\end{algorithm}

The constructed S-DAGs serve as supervision signals for training the GNN module. This enables the model to learn to predict relevant subjects and their dependencies during inference. The full training procedure is detailed in Alg.~\ref{alg:gnn_training}.

\section{Multi-Agent Collaborative Inference}

\label{app:colaboraive-inference}

\subsection{Model Pool}
\label{app:model-pool}

Given the subject decomposition provided by the constructed S-DAG, we are able to precisely identify the domain expertise required to solve each question. This enables targeted selection of specialized expert agents—language models fine-tuned for specific subjects.

To support this, we curate a Model Pool consisting of small yet high-performing subject-specialized LLMs, sourced primarily from the Hugging Face model repository. Each model is carefully chosen based on its training corpus, fine-tuning objectives, and performance within its respective domain. Through the LLM subject capability profiling process, we can conduct precise subject-LLM matching, ensuring that each subject node in the S-DAG is handled by an appropriately skilled agent.

The full list of expert models, along with their associated subjects and Hugging Face links, is presented in Table~\ref{tab:model-list}. These models collectively enable a modular, scalable approach to complex multi-subject reasoning, offering both computational efficiency and domain fidelity.

\begin{table*}[t!]
\centering
\small
\caption{Subject-specific base models with hugging face links.}
\label{tab:model-list}
\scalebox{0.9}{
\begin{tabular}{lll}
\toprule
\textbf{Subject} & \textbf{Base Model} & \textbf{Link} \\
\midrule
Chemistry & Llama2-13B-Chat & https://huggingface.co/juntaoyuan/chemistry-assistant-13b \\
Math & Deepseek-7B & https://huggingface.co/deepseek-ai/deepseek-math-7b-instruct \\
Biomedical & Llama-3-8B & https://huggingface.co/ContactDoctor/Bio-Medical-Llama-3-8B \\
Biology & BioMistral-7B & https://huggingface.co/BioMistral/BioMistral-7B \\
Computer Science & Qwen2.5-7B & https://huggingface.co/Qwen/Qwen2.5-Coder-7B-Instruct \\
Physician & Llama-3-8B & https://huggingface.co/YiDuo1999/Llama-3-Physician-8B-Instruct \\
Law & Llama3-8B & https://huggingface.co/ricdomolm/lawma-8b \\
Economics & Mistral-7B & https://huggingface.co/tim9510019/Mistral-7B-Economic\_zephyr\_231023 \\
Business & Llama-3.1-8B & https://huggingface.co/warrencain/Business\_Consulting\_Finetune\_Llama\_3.1\_8b \\
History & Deepseek-R1-distill-Qwen-7B & \text{https://huggingface.co/Doppelfelix/Deepseek-r1-history-expert} \\
Engineering & Llama2-7B-Chat & https://huggingface.co/Frrrrrrrrank/Llama-2-7b-chat-hf-process\_engineering\_one\_firsttwokap\_v3 \\
Psychology & Llama-3.1-8B & https://huggingface.co/dentist9111/Llama-3.1-8B-bnb-4bit-samhog\_psychology \\
Philosophy & Mistral-12B & https://huggingface.co/EpistemeAI/Mistral-Nemo-Instruct-12B-Philosophy-Math \\
Health & Llama3-8B & https://huggingface.co/m42-health/Llama3-Med42-8B \\
\bottomrule
\end{tabular}}
\end{table*}

\begin{figure}[t!]
    \small
    \centering
    \includegraphics[width=0.5\linewidth]{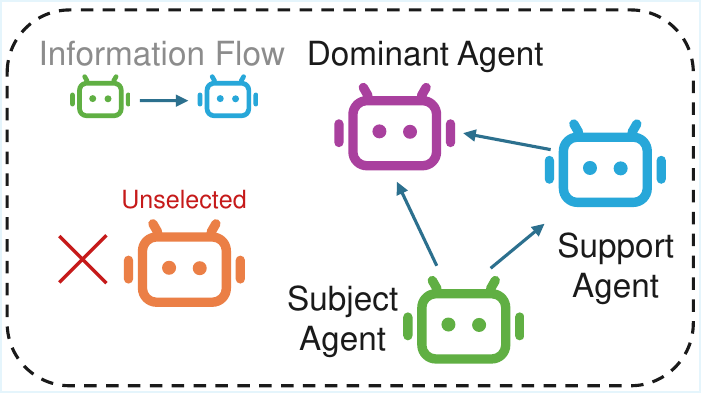}
    \caption{The S-DAG structured multi-agent information flow.}
    \label{fig:info-flow}
\end{figure}

\subsection{Multi-Agent Information Flow}
\label{app:information-flow}

Once the subject-directed acyclic graph is constructed from a heterogeneous question, we can proceed to the subject–LLM pairing phase based on the LLM subject capability profiles. In this step, each node in the S-DAG—representing a specific subject—is assigned to a domain-specialized expert LLM, enabling targeted reasoning within that domain.

We define three types of agents based on their structural position in the S-DAG:

\begin{itemize}
    \item \textbf{Subject Expert Agents}: These are agents corresponding to the starting nodes, which initiate the reasoning process. They do not receive input from other agents but process the raw question directly within their area of expertise.

    \item \textbf{Supporting Agents}: Associated with intermediate nodes, these agents integrate information received from upstream agents and generate enriched outputs. They act as conduits, facilitating inter-subject reasoning and knowledge flow across the S-DAG.

    \item \textbf{Dominant Agent}: This agent corresponds to the final node in the graph. It receives input from multiple supporting agents but does not transmit information further. The Main Agent is responsible for producing the final answer, synthesizing interdisciplinary insights gathered throughout the reasoning chain.

\end{itemize}

All agents receive the original question and operate collaboratively within the information flow defined by the S-DAG. Their internal processing and response strategies are guided by the structural role they play—whether initiating, transforming, or concluding the reasoning. This multi-agent orchestration enables both modular specialization and coherent integration across subject domains, reflecting the layered nature of complex human reasoning. Figure \ref{fig:info-flow} shows the multi-agent information flow under S-DAG. Table \ref{tab:pmt-info-flow} shows the prompts for different types of agents.

\begin{table*}[t!]
\small
\caption{Prompts for S-DAG structured multi-agent information flow.}
\label{tab:pmt-info-flow}
\centering
\begin{tabular}{p{0.25\linewidth}p{0.7\linewidth}}
\toprule
\textbf{Agent} & \textbf{Prompt} \\
\midrule
\textbf{Subject Expert Agent} & \multicolumn{1}{m{0.7\linewidth}}{
You are an expert in [Subject]. Your task is to analyze the following question based on your domain knowledge.
Question: \{Q\}
Please provide a clear and concise explanation or answer strictly from the perspective of [Subject].} \\
\midrule
\textbf{Supporting Agent} & \multicolumn{1}{m{0.7\linewidth}}{You are an expert in [Subject]. Another agent has provided information from [Supporting Subject], which may be relevant to your reasoning.
Question: \{Q\}
Supporting Information from [Supporting Subject]: \{Support Content\}
Please incorporate the above supporting information into your domain-specific reasoning, and produce a coherent, informed response from the perspective of [Subject].} \\
\midrule
\textbf{Dominant Agent} & \multicolumn{1}{m{0.7\linewidth}}{You are the lead [Subject] expert responsible for integrating multi-disciplinary information to answer the following complex question.
Question: \{Q\}
You have received input from other experts:
- [Subject A]: \{Content A\}
- [Subject B]: \{Content B\}
\ldots
Please synthesize the provided information and generate a comprehensive final answer that reflects the reasoning across these domains.} \\
\bottomrule
\end{tabular}
\end{table*}

\begin{table*}[t!]
\small
\caption{Prompts for baselines. List of prompts used in experiment.}
\label{tab:pmt-for-baseline}
\centering
\begin{tabular}{p{0.25\linewidth}p{0.7\linewidth}}
\toprule
\textbf{Baseline} & \textbf{Prompt} \\
\midrule
\textbf{Single-Model Method} & \multicolumn{1}{m{0.7\linewidth}}{Can you solve the problem? \{Q\} Explain your reasoning. Your final answer should be with the format: \textless{}\textless{}answer\textgreater{}\textgreater{}, at the end of your response.} \\ \midrule
\textbf{Self-Refine} & \multicolumn{1}{m{0.7\linewidth}}{1. Initial Answer: Can you solve the problem? \{Q\} Explain your reasoning. Your final answer should be with the format: \textless{}\textless{}answer\textgreater{}\textgreater{}, at the end of your response. 2. Feedback: The question is: \{Q\} Please identify issues or limitations in the following answer and suggest improvements: \{answer\} 3. Refine: The problem is: \{Q\} The original answer was: \{answer\} The feedback is:\textbackslash{}n\{feedback\_content\}\textbackslash{}n\textbackslash{}nNow give an improved correct answer based on this feedback. Your final answer should be with the format: \textless{}\textless{}answer\textgreater{}\textgreater{}, at the end of your response.} \\ \midrule
\textbf{Multi-Agent Debate} & \multicolumn{1}{m{0.7\linewidth}}{1. Statement: You are Agent [A]. Can you solve the problem? {Q} Explain your reasoning. 2. Rebuttal: You are Agent [A]. Rebut Agent [B]'s argument below: Agent [B] said: \{response\} 3. Judge: You are the Judge. Based on both arguments and rebuttals, decide which Agent has a stronger case. Agent [A] said: \{response\} Agent [B] said: \{response\}}\\

\bottomrule
\end{tabular}
\end{table*}

\section{Ablation Study}
\label{app:ablation}

To better understand the contribution of each component in our framework, we conduct a series of ablation experiments on the MMLU-Pro, GPQA, and MedMCQA benchmarks. Specifically, we evaluate the effectiveness of the GNN module for S-DAG generation, the LLM subject capability profiling for model selection, and the directed acyclic graph topology itself. Results are summarized in Table~\ref{tab:ablation}.

To evaluate efficiency, we report the average inference time and number of LLM calls per instance across the test set. For each sample, inference time is measured from the moment the question is input into the system until the final answer is generated, including all model invocation, data routing, and inter-agent communication. The number of LLM calls is computed by counting each individual model invocation used to process subject nodes within the S-DAG. For example, if three subject nodes are activated and each requires a single forward pass through its assigned expert model, this counts as three LLM calls. The final reported values are averaged over all test questions to reflect overall system efficiency. All experiments are conducted under controlled conditions on the same hardware to ensure fair comparison.

\subsection{GNN Module}
The GNN module plays a central role in generating the S-DAG, which guides multi-agent collaboration. While it is possible to rely solely on LLMs to extract subject annotations and manually construct a DAG, this approach is inefficient and error-prone. LLM outputs tend to be noisy, inconsistent across runs, and may lack a coherent structure. 

In contrast, our trained GNN is able to learn subject relevance and interdependencies across a large number of annotated samples, yielding robust and repeatable DAGs. To evaluate the GNN's contribution, we replace it with an LLM-based rule system that directly constructs the S-DAG using raw subject weights from the LLM.

\subsection{LLM Profile}
The LLM profile component enables accurate subject–model matching by assigning each expert model a capability score for each subject domain. Without this profiling mechanism, subject nodes in the S-DAG are matched to expert LLMs randomly, ignoring domain expertise. 

To measure the impact of profiling, we compare our system against a variant where expert agents are selected uniformly at random for each subject. This leads to inconsistent or suboptimal assignments, resulting in a noticeable performance drop. The ablation confirms that subject-aware model selection is critical for effective multi-agent collaboration, as it ensures that each subtask is handled by the most capable model based on prior subject-specific evaluations.

\subsection{Graph Topology}
Our framework relies on a directed acyclic graph (DAG) structure to represent the subject-level reasoning flow, enabling information to propagate from supporting subjects to dominant ones in a structured manner. To assess the importance of this topology, we compare the directed S-DAG against a fully connected subject graph, where all subject nodes can freely exchange information without structural constraints.

Although the fully connected variant allows maximum communication among agents, it introduces redundant interactions and potential reasoning loops, increasing inference cost and ambiguity. The directed version not only improves accuracy but also reduces inference time and LLM calls. These results highlight the importance of enforcing a sparse and efficient graph structure.

\section{Broader Impacts}
\label{app:broader-impact}
Our work introduces S-DAG, a modular framework for multi-subject reasoning that leverages lightweight, domain-specific language models. By enabling more efficient and scalable deployment compared to large monolithic LLMs, S-DAG holds promise for democratizing access to advanced AI systems, particularly in education, healthcare, and scientific research. Its graph-structured design promotes transparency and modularity, supporting safer and more controllable reasoning pipelines. However, potential risks must also be acknowledged. The framework relies on subject-specific models that may vary in quality or harbor domain-specific biases, which could propagate through the reasoning process. Furthermore, improper deployment in high-stakes applications without adequate validation may lead to unintended consequences. Future work should explore robust evaluation standards, fairness auditing, and safeguards to ensure responsible use of S-DAG in real-world scenarios.

\end{document}